\documentclass[a4paper,11pt]{article}
\usepackage{pos}
\usepackage{bibspacing}

\newcommand{\LQCD}{\Lambda_{\rm QCD}}

\newcommand{\DY}{\Delta Y}

\newcommand{\tcite}[1]{~\cite{#1}}
\newcommand{\tref}[1]{~\ref{#1}}
\newcommand{\eref}[1]{~\eqref{#1}}
\newcommand{\drv}{{\rm d}}

\title{Higgs-plus-jet inclusive production as stabilizer of the high-energy resummation}
\ShortTitle{Higgs-plus-jet differential distributions}

\author*[a,b,c]{Francesco Giovanni Celiberto}
\author[d,e]{Michael Fucilla}
\author[f]{Dmitry Yu. Ivanov}
\author[d,e]{Mohammed M.A. Mohammed}
\author[d,e]{Alessandro Papa}

\affiliation[a]{European Centre for Theoretical Studies in Nuclear Physics and Related Areas (ECT*), I-38123 Villazzano, Trento, Italy}
\affiliation[b]{Fondazione Bruno Kessler (FBK), I-38123 Povo, Trento, Italy}
\affiliation[c]{INFN-TIFPA Trento Institute of Fundamental Physics and Applications, I-38123 Povo, Trento, Italy}
\affiliation[d]{Dipartimento di Fisica, Universit\`a della Calabria, I-87036 Arcavacata di Rende, Cosenza, Italy}
\affiliation[e]{Istituto Nazionale di Fisica Nucleare, Gruppo collegato di Cosenza I-87036 Arcavacata di Rende, Cosenza, Italy}
\affiliation[f]{Sobolev Institute of Mathematics, 630090 Novosibirsk, Russia}

\emailAdd{fceliberto@ectstar.eu}
\emailAdd{michael.fucilla@unical.it}
\emailAdd{d-ivanov@math.nsc.ru}
\emailAdd{mohammed.maher@unical.it}
\emailAdd{alessandro.papa@fis.unical.it}

\abstract{
We investigate the inclusive hadroproduction of a Higgs boson in association with a jet, featuring large transverse momenta and separated by a large rapidity distance. We propose this reaction, that can be studied at the LHC as well as at new-generation colliding machines, as a novel probe channel for the manifestation of the Balitsky--Fadin--Kuraev--Lipatov (BFKL) dynamics. We bring evidence that high-energy resummed distributions in rapidity and transverse momentum exhibit a solid stability under higher-order corrections,
thus offering us a faultless chance to gauge the feasibility of precision calculations of these observables at high energies. We come out with the message that future, exhaustive analyses of the inclusive Higgs-boson production, would benefit from the inclusion of high-energy effects in a \emph{multi-lateral} formalism where distinct resummations are concurrently embodied.
We propose these studies with the aim of inspiring synergies with other Communities, and pursuing the goal of widening common horizons in the exploration of the Higgs-physics sector.
}

\FullConference{%
  *** The European Physical Society Conference on High Energy Physics (EPS-HEP2021), ***\\
  *** 26-30 July 2021 ***\\
  *** Online conference, jointly organized by Universität Hamburg and the research center DESY ***
}

\begin{document}
\maketitle

\section{Hors d'{\oe}uvre}
\label{introduction}

The Higgs sector is universally recognized as one of the golden channels where to hunt for long-awaited signals of New Physics.
At the same time, Higgs emissions in forward directions of rapidities offer us a unique chance to probe kinematic corners where our knowledge of the Standard Model, although being well established, needs to be further expanded.
One of these corners is represented by the so-called \emph{semi-hard} regime, namely where final-state configurations provide us with a stringent scale ordering, $\sqrt{s} \gg \{Q\} \gg \LQCD$. Here, $\sqrt{s}$~is the center-of-mass energy, $\{Q\}$ one or a set of hard scales given by the kinematics, and $\LQCD$ the QCD scale.
In this regime, genuine fixed-order calculation within perturbative QCD would fail, since large energy logarithms enter the perturbative series with a power that increases with the perturbative order, thus systematically compensating the narrowness of the strong coupling constant. 
The Balitsky--Fadin--Kuraev--Lipatov (BFKL) approach~\cite{Fadin:1975cb,Kuraev:1976ge,Kuraev:1977fs,Balitsky:1978ic} allows us to resum to all orders these large logarithms, up to the leading (LLA) and next-to-leading logarithmic approximation (NLA).
In this framework, cross sections for hadronic processes are written as a convolution of two process-dependent impact factors, depicting the fragmentation of each colliding particle to an identified final-state object, and a universal Green's function, whose evolution kernel is known within next-to-leading (NLO) accuracy.
In the last years a relevant number of semi-hard reactions was proposed, that can be studied via proton-proton collisions at the LHC. An incomplete list includes: the inclusive hadroproduction two Mueller--Navelet jets\tcite{Mueller:1986ey}, for which several analyses have been conducted so far~\tcite{Colferai:2010wu,Caporale:2012ih,Ducloue:2013bva,Caporale:2014gpa,Celiberto:2015yba,Celiberto:2015mpa,Celiberto:2016ygs,Celiberto:2016vva,Caporale:2018qnm}, the inclusive light hadron production\tcite{Celiberto:2016hae,Celiberto:2016zgb,Celiberto:2017ptm,Bolognino:2018oth,Bolognino:2019yqj,Bolognino:2019cac,Celiberto:2020rxb}, multi-jet production\tcite{Caporale:2015vya,Caporale:2015int,Caporale:2016soq,Caporale:2016vxt,Caporale:2016xku,Celiberto:2016vhn,Caporale:2016djm,Caporale:2016lnh,Caporale:2016zkc}, $J/\psi$-jet\tcite{Boussarie:2017oae,Celiberto:2022dyf}, heavy-flavor\tcite{Bolognino:2019ouc,Bolognino:2019yls,Bolognino:2021mrc,Bolognino:2021hxx,Celiberto:2021dzy,Celiberto:2021fdp,Bolognino:2022wgl}, and forward Drell–Yan plus jet\tcite{Golec-Biernat:2018kem} correlations.
In this work we propose a novel semi-hard channel, \emph{i.e.} the inclusive detection of a Higgs boson in association with a light jet, both of them emitted with high transverse momentum and separated by a large rapidity interval.
Here, by virtue of the large transverse masses provided by Higgs emissions, clear signals of a reached stability of the high-energy resummed series under higher-order corrections and scale variation are expected.

\section{Inclusive Higgs-plus-jet production at the LHC}
\label{results}

\begin{figure}[t]

   \includegraphics[scale=0.49,clip]{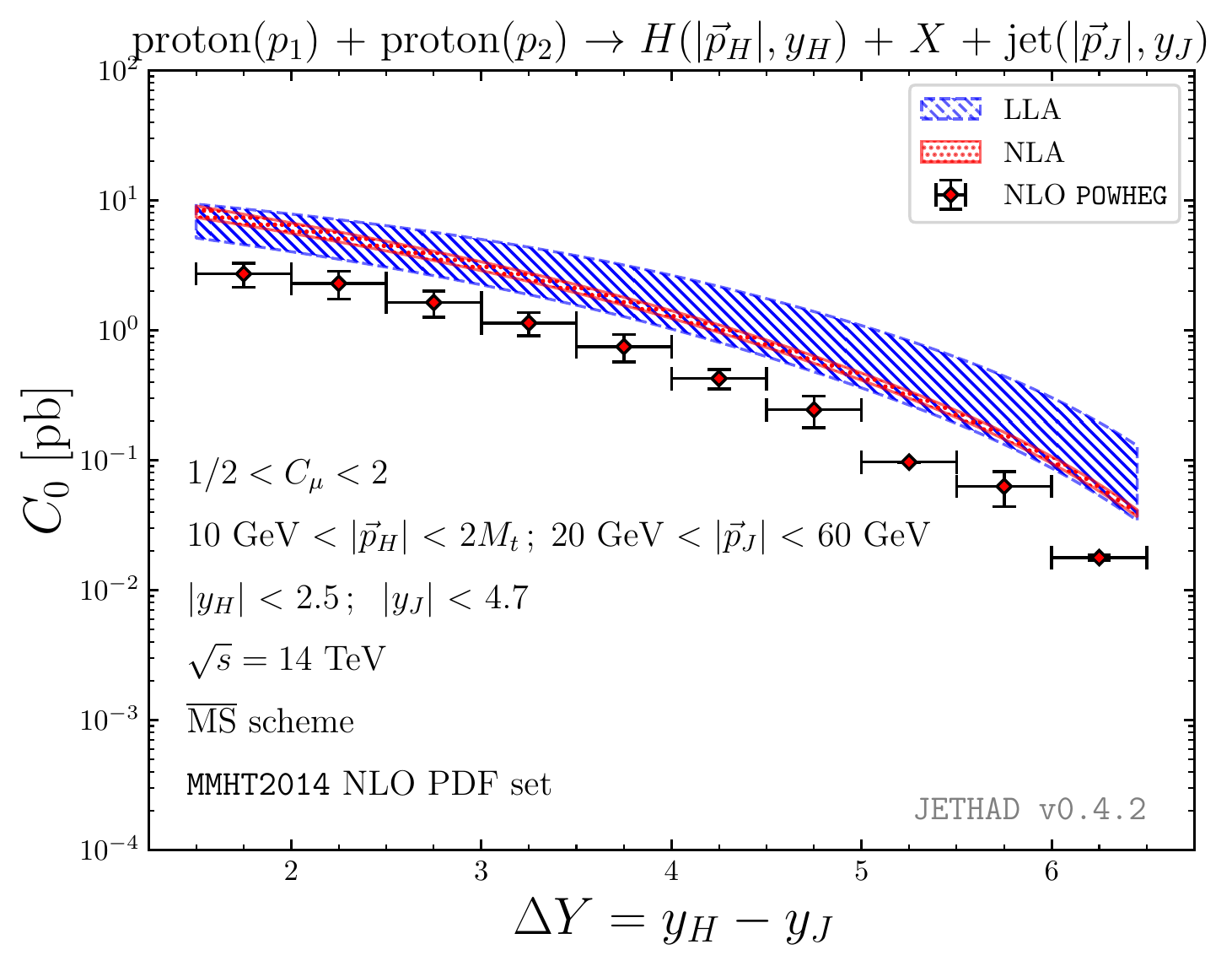}
   \includegraphics[scale=0.49,clip]{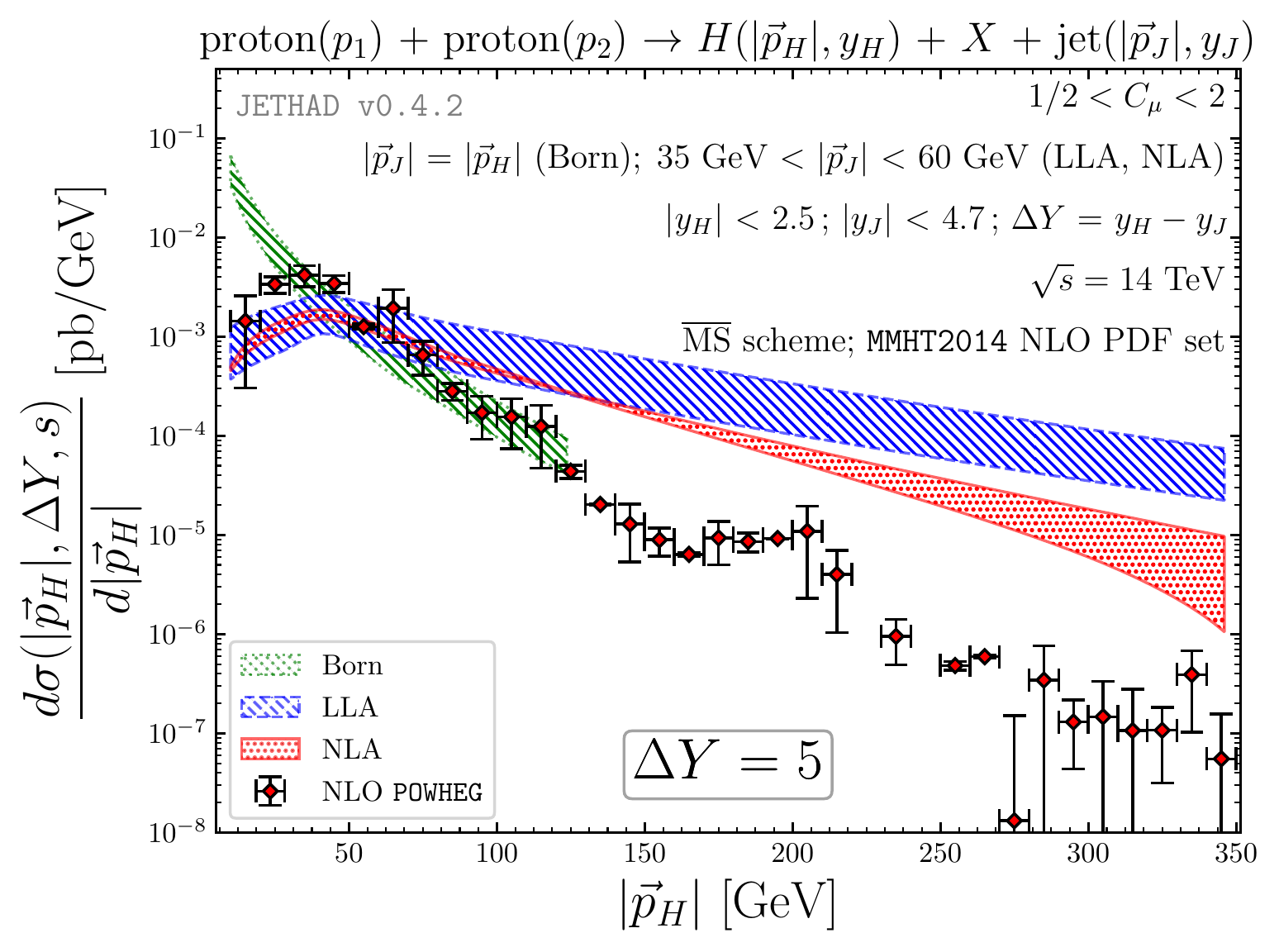}

\caption{$\DY$-shape of the $C_0$ (left) and $|\vec p_H|$-behavior of the Higgs transverse-momentum distribution at $\DY = 5$ (right). Text boxes inside panels show transverse-momentum and rapidity ranges. Uncertainty bands embody the combined effect of scale variation and phase-space multi-dimensional integration.}
\label{fig:distributions}
\end{figure}

We consider the inclusive hadroproduction of a Higgs boson and a jet,
\begin{equation}
\label{process}
 {\rm proton}(p_1) \, + \, {\rm proton}(p_2) \, \to \, H(|\vec p_H|, y_H) \, + \, {\cal X} \, + \, {\rm jet}(|\vec p_J|, y_J) \;,
\end{equation}
where ${\cal X}$ stands for an undetected final-state gluon system.
The Higgs and the jet are emitted with large transverse momenta, $|\vec p_{H,J}|$, and separated by a large rapidity distance, $\Delta Y \equiv y_H - y_J$. It is convenient to represent the high-energy resummed cross section as a Fourier sum of the azimuthal-angle coefficients
\begin{equation}
\label{sigma_Fourier}
 \frac{\drv \sigma}{\drv y_H \drv y_J \drv |\vec p_H| \drv |\vec p_J| \drv \phi_H \drv \phi_J} = \frac{1}{4 \pi^2} \left[ {\cal C}_0 + \sum_{k=1}^{+ \infty} 2 \cos(k \varphi) {\cal C}_k \right] \;,
\end{equation}
with $\phi_{H,J}$ the Higgs (jet) azimuthal angles and $\varphi \equiv \phi_H - \phi_J - \pi$. In Eq.\eref{sigma_Fourier} ${\cal C}_0$ is the $\varphi$-summed cross section and $C_{k > 0}$ determine the final-state azimuthal distribution.
Starting from our fully differential cross section, we consider two observables that can be studied at current LHC kinematic configurations. The first one is the $\DY$-distribution (or simply $C_0$)
\begin{equation}
\label{C0}
 C_0(s, \Delta Y) = 
 \int_{y_H^{\rm min}}^{y_H^{\rm max}} \drv y_H
 \int_{y_J^{\rm min}}^{y_J^{\rm max}} \drv y_J
 \int_{p_H^{\rm min}}^{p_H^{\rm max}} \drv |\vec p_H|
 \int_{p_J^{\rm min}}^{p_J^{\rm max}} \drv |\vec p_J|
 \, \,
 \delta (\DY - (y_H - y_J))
 \, \,
 {\cal C}_0
 \, ,
\end{equation}
and the second one is the Higgs transverse-momentum distribution (or simply $|\vec{p_H}|$-distribution)
\begin{equation}
\label{pH}
 \frac{\drv \sigma(|\vec p_H|, s, \Delta Y)}{\drv |\vec p_H| \drv \DY} = 
 \int_{y_H^{\rm min}}^{y_H^{\rm max}} \drv y_H
 \int_{y_J^{\rm min}}^{y_J^{\rm max}} \drv y_J
 \int_{p_J^{\rm min}}^{p_J^{\rm max}} \drv |\vec p_J|
 \, \,
 \delta (\DY - (y_H - y_J))
 \, \,
 {\cal C}_0
 \, .
\end{equation}
For both the observables we constrain the Higgs emission inside rapidity acceptances of the CMS barrel detector, $|y_H| < 2.5$, while we allow for a larger rapidity range of the light jet, that can be detected also by the CMS endcaps, $|y_J| < 4.7$. We consider an \emph{asymmetric} configuration of $p_T$-ranges for $C_0$, namely $10 < |\vec p_H|/{\rm GeV} < 2 m_t$ ($m_t$ is the top-quark mass) and $20 < |\vec p_J|/{\rm GeV} < 60$, while we set $35 < |\vec p_J|/{\rm GeV} < 60$ for the $|\vec{p_H}|$-distribution. The center-of-mass energy squared is fixed at $\sqrt{s} = 14$~TeV. For a detailed study on different kinematic ranges, see Refs.\tcite{Celiberto:2020tmb,Celiberto:2021fjf,Celiberto:2021txb}. 
In Fig.\tref{fig:distributions} we compare the LLA- and NLA-resummed behavior of our distributions, calculated via the {\tt JETHAD} modular work package\tcite{Celiberto:2020wpk}, with NLO fixed-order predictions obtained through the {\tt POWHEG} method\tcite{Frixione:2007vw,Alioli:2010xd,Campbell:2012am}.
We observe that NLA predictions (red) for $C_0$ (left panel) are almost completely nested inside LLA uncertainty bands (blue), thus corroborating the underlying assumption that large energy scales afforded by the Higgs-boson detection act as \emph{stabilizers} for the BFKL series.
This feature emerges also for the $|\vec p_H|$-distribution (right panel) in the intermediate-$|\vec p_H|$ region, namely the peak region plus the first part of the decreasing tail, where NLA bands are totally contained inside the LLA ones. Here, the impressive stability of the resummed series unambiguously confirms the validity of our description by the hand of the BFKL approach. However, in the large-$|\vec p_H|$ region by the long
tail, NLA predictions decouple from LLA results and show an increasing sensitivity to energy-scale variation. Here, DGLAP-type logarithms together with \emph{threshold} effects become more and more significant, thus worsening the convergence of the high-energy series.
Finally, we notice that NLO fixed-order results are systematically lower than high-energy resummed ones and this is more evident at the larger $\DY$-values, where resummation effects are expected to be relevant.

\section{Toward new directions}
\label{conslusions}

We have proposed the inclusive semi-hard production of a Higgs-plus-jet system in hadronic collisions as a novel channel for the manifestation of stabilization effects of the high-energy resummation under higher-order corrections.
We discovered that those effects are present and allow for the description of distributions differential in rapidity and transverse momentum at natural scales.
Statistics for cross sections, shaped on realistic LHC kinematic configurations, is encouraging.
We propose, as a medium-term outlook, two new directions.
On the one hand, the study of the single forward or central inclusive emission of a Higgs boson in proton-proton scatterings represents a clear channel to access the proton structure at low-$x$. More in particular, analyses on Higgs production in the aforementioned rapidity configurations will sensibly extend our knowledge of the BFKL \emph{unintegrated gluon distribution} (UGD) in the proton, thus enriching current phenomenological studies done via forward vector-meson\tcite{Anikin:2011sa,Besse:2013muy,Bolognino:2018rhb,Bolognino:2018mlw,Bolognino:2019bko,Bolognino:2019pba,Celiberto:2019slj,Bautista:2016xnp,Garcia:2019tne,Bolognino:2021niq,Bolognino:2021gjm,Bolognino:2022uty} and forward Drell--Yan dilepton pair\tcite{Brzeminski:2016lwh,Celiberto:2018muu} emissions, and possibly shedding light on common ground between BFKL and other formalisms where small-$x$ gluon densities are defined, such as the transverse-momentum-dependent (TMD) factorization (see Refs.\tcite{Bacchetta:2020vty,Celiberto:2021zww,Celiberto:2022fam,Bacchetta:2021oht,Bacchetta:2021lvw,Bacchetta:2021twk,Bacchetta:2022esb} for recent applications on gluon dynamics) and Altarelli--Ball--Forte (ABF) resummed collinear distributions\tcite{Altarelli:2001ji,Ball:2017otu,Abdolmaleki:2018jln}.
On the other hand, the fair stability of our distributions motivates our interest evolving our formalism into a \emph{multi-lateral} approach that embodies different resummations, to be tested at current and new-generation colliding facilities\tcite{AbdulKhalek:2021gbh,Arbuzov:2020cqg,Chapon:2020heu,Anchordoqui:2021ghd}.

\vspace{-0.25cm}
\bibliographystyle{bibstyle}
\bibliography{bibliography}

\end{document}